# Loops and Knots as Topoi of Substance
## Spinoza Revisited[1]

**Rainer E. Zimmermann**

IAG Philosophische Grundlagenprobleme
FB 1, UGH, Nora-Platiel-Str.1, D - 34127 Kassel
Clare Hall, UK - Cambridge CB3 9AL[2]
Lehrgebiet Philosophie
FB 13 AW, FH, Lothstr.34, D - 80335 Muenchen[3]
pd00108@mail.lrz-muenchen.de

April 27, 2000
gr-qc/0004077 v2 (May 22, 2000)

## Abstract

The relationship between modern philosophy and physics is discussed. It is shown that the latter develops some need for a modernized metaphysics which shows up as an *ultima philosophia* of considerable heuristic value, rather than as the *prima philosophia* in the Aristotelian sense as it had been intended, in the first place. It is shown then, that it is the philosophy of Spinoza in fact, that can still serve as a paradigm for such an approach. In particular, Spinoza's concept of infinite substance is compared with the philosophical implications of the foundational aspects of modern physical theory. Various connotations of substance are discussed within pre-geometric theories, especially with a view to the role of spin networks within quantum gravity. It is found to be useful to introduce a separation into physics then, so as to differ between foundational and empirical theories, respectively. This leads to a straightforward connection between foundational theories and speculative philosophy on the one hand, and between empirical theories and sceptical philosophy on the other. This might help in the end, to clarify some recent problems, such as the absence of time and causality at a fundamental level. It is implied that recent results relating to topos theory might open the way towards eventually deriving logic from physics, and also towards a possible transition from logic to hermeneutic.

---

[1] Parts of this paper have been presented in two seminar talks given in the Michaelmas term 1999 at the History and Philosophy of Science Department, Cambridge (UK), and in a talk to the Physics Theory Group at Birkbeck College, London, in the following December. A revised version of this paper will be published under the title „Spinoza in Context. A Holistic Approach in Modern Terms." as an invited contribution to E.Martikainen (ed.), Infinity, Causality, and Determinism, Cosmological Enterprises and their Preconditions, Helsinki, Finland, 8-10 May 2000.
[2] Permanent addresses.
[3] Present address.

# 1 Philosophy and Science Today

The main task of humans is to model nature. To be more precise, they model their surrounding environment, and they model themselves within that environment. This serves as a guideline for orientation within a complex world and gives them conceptions of how to actually deal with this world in practise. From what they have found out about the environment in their immediate vicinity, they infer results about phenomena and things which are far away (both in spatial and temporal terms, respectively). The idea is to generalize the results as far as possible; this stabilizes experience in terms of the modelling done. Stable (in fact, *structurally* stable) aspects of nature can be taught to others, because in that case the models achieved imply general rules and a systematic approach to nature which can be communicated easily. In principle then, teaching is „saving experience" (if only for relatively simple things such as the sciences).

At the same time, humans are also a *product* of nature. That is, they have emerged from the same process they are trying to model, in first place. By having been produced, they now interact with nature in order to actively modify it according to their objectives. They are participators rather than mere observers. In so far as they are nature's product, their modifying of nature is still the same process representing nature's activity. Hence, in principle, nature is actually modifying *itself* by means of human (research) activities.

Obviously, the various fields of scientific activity are divided according to the „regional" aspects nature is displaying for the various perspectives of systematic interest. The division of labour is also applied to „research labour" in the sense that it reduces the observed complexity of the world. In fact, thinking itself is deeply rooted in this permanent activity of reducing complexity (which could be used as another formula for expressing the modelling activity of humans). Humans think in a „digital" rather than analogous way, that is they artificially introduce sections and intervals, or distances into the process they actually observe, they „sequentialize" this process, which remains however all the time quite an analogous (continuous) process, if viewed in terms of perception (even if its structure is highly non-linear). On the other hand, modern physics tends to show that space is intrinsically *digitalized* in a way, and so might be time (if it is not absent altogether). And this result would not come as a surprise, in fact, because of humans being a product of nature, leading to the conclusion that their models of nature become more adapted to their (natural) mode of thinking in the long run. Hence, the contradiction between continuous and discrete structures of the world (leading to different concepts of process) has always been very much in the centre of discussion. Although the modelling performed introduces a level of abstraction into nature which apparently leads away from concrete, practical nature, it is nevertheless nearer to reality in the sense that humans may learn to differentiate between how the world appears to be (in its modality) and how it may be in reality. In other words, humans perceive the world according to their biological constitution which is an outcome of the process of nature. And this constitution comprises e.g. of temporality. But these means of perception opening up the way to cognitively grasp the world, are aimed towards optimizing orientation within the world (for finding enough food and being sufficiently protected in the first place). There is no reason to believe that there is any necessity involved in a fitting of these perceptive capacities to the actual uncovering of the real structure of the world. Without doubt, there *is* some real world of which humans are a part, and obviously, it *can* be mapped adequately in one way or the other, but this does not tell us any kind of absolute truth about the being of the world. In a sense, the speculative abilities of humans can be visualized as a kind of excess knowledge achieved (probably because the impetus of striving for insight has a kind of dynamical space of free play which within the field of its anticipations opens up considerably more possibilities than are actually needed for the time being.) The *topics* of speculation however, as well as the topics of empirical research have not really changed, because, given the human relation to environment, the choice is probably limited.

The traditional division of scientific subjects, having actually been developed by stoic philosophers for the first time (within European thought) in all generality, has remained essentially invariant until today, though the terminology and the conceptual attitude might have changed considerably. In this sense, *physics* is dealing with the modelling of nature, *logic* with the modelling of thinking about (modelling) nature, and *hermeneutic* with the modelling of thinking about thinking. (I do impose here a mild reductionism in using epistemological core categories rather than fields of research. In this sense, the sciences are *reducible* to physics such that the latter is included in them as a special case at their basic foundations. But higher levels of complexity are *irreducible* to physics in the sense that by passing down the hierarchy, emergent properties are effectively being lost.[4]) And we can say something more about this: Because of humans being a product of nature, their models of nature must necessarily contain *self-models*. In fact, it may be possible to formulate (and prove) a simple proposition stating that each model of nature contains at least one self-model.[5]

---

[4] This is an almost trivial point which coincides with the traditional view of what is generally called „dialectic materialism." Unfortunately, in ongoing debates, „reductionsm" is mostly visualized instead in terms of „mechanical materialism" in the sense of 19th century views or misled ideological arguments put forward in the earlier epochs of the former Soviet Union.

[5] This is in fact an almost trivial proposition again. I have explained this in more detail in my paper „Ordnung und Unordnung - Zum neueren Determinismusstreit zwischen Thom und Prigogine", Lendemains 50, Berlin,

So we may come to the conclusion that all this modelling going on is primarily one of a unified character, because it is dealing with a unified whole, a totality which cannot actually be broken up into parts, if not accepting that important information will be lost then. On the other hand, because of practical reasons, it is not possible to model everything at the same time. Hence, humans have to take into account both these aspects: They have to study the minute details of nature, but they have also to think about the interactive pattern which is constituting nature's totality, of which they are actually a part themselves. For the one task they do have the sciences which deal with a fragment of nature only, according to the definitions of their „regional competence", for which they are designed in first place, and according to their prescribed boundaries (which are sometimes rather fluid nowadays). The problem with this approach is that very often, research in one field is not at all related to (or even knows of) research in another so that the interrelationship of these fragments forming a whole is more or less lost. That is why philosophy takes the other perspective and centres around the aspect of totality. But in doing that, nowadays, philosophy has developed as a science of sciences, or as a science of totality, following up the scientific results given at a time, rather than trying to lay the grounds for science, as it has been the tradition for a long time. Contrary to what was intended by the Aristotelian idea of „prima philosophia", philosophy is now putting forward research of a more conceptual and heuristic kind.[6]

What we will do in the following is to recall basic aspects of Spinoza's concept of infinite substance (section II) in order to illustrate one of the chief origins of the view presented here. The concept of substance proper is discussed then in more detail, as a paradigm for basic ideas which are still relevant for us today (section III). Recent foundational developments in physics are shortly summarized, and their relationship to philosophical connotations of substance metaphysics is shown, with a special stress being laid upon pre-geometric aspects of emergence (section IV). Topos-theoretic consequences are discussed then (section V), which might open the perspective towards an explicit derivation of logic from physics (with obvious materialistic implications). Preliminary conclusions are finally drawn (section VI) as to a possible transgression of these results towards a transition from logic to hermeneutic.

As might be appropriate for an essentially interdisciplinary „joint venture", both the philosophical sections (II, III), and the physical and mathematical sections (IV, V), have been written on what the author thought would be a medium level of technicality. Hence, the former are composed in a somewhat abstract, but at the same time basically narrative language. The latter shortly summarize the tech-nical aspects being accompanied by passages of philosophical commentary. The important point is to be able to re-locate the relevant aspects of the former in the latter, and also viceversa, and to make possible an interdisciplinary „cross-reading". As this is not very easy to achieve, some detailed explanations of a conceptual nature have been included in this paper, which also accounts for its length.

## 2 Spinoza Revisited

Basically, all of this is not really a new idea: It was already Spinoza in fact, who had thought about these aspects in systematic detail. (Although this line of thought can also be found much earlier in stoic philosophy, as mentioned above.) As I have shown at another place[7], for a long time, the consequent resumption and continuation of the stoic line was obstructed due to ideological reasons, especially with a view to the dynamical aspects of nature primarily based on an active sort of „creative matter" leading forward to the possibility of a field concept with regard to physical forces. Ironically however, it was exactly this sort of negative approach which - as a kind of by-product - made it possible to gain more insight into the relationship between the world and its foundation in the long run, because the contradictions encountered sooner or later did not leave another way out than just the resumption of the stoic line. The permanent spinning around the relationship between God and world ended soon in blind alleys and on wrong tracks. Their by-passing in terms of modern (Renaissance) mathematics opened up an alternative approach with a two-fold result: On the one hand, the ancient concept of substance was re-formulated in an abstract way, in explicitly denying its existence in the traditional, scholastic sense, thus re-introducing it as a concept in terms of a productive misunderstanding and in a somewhat clandestine way. On the other hand, the progressing detachment of philosophy from theology opened up a new way to all aspects of nature which had been suppressed before: such as the dynamics of self-creative and self-organizing matter as substratum and potential of an unfolding nature, including a re-definition of the organizational hierarchy of the universe, and of the mediation of matter with its explicit forms up to its social forms constituting human systems, themselves dealing with the reflexion of this very process, being subjected however to incomplete information. So it was left to Spinoza to actually draw all these aspects (which were drifting apart for quite a while) together again and to re-base them onto a sound foundation of stoic systematics, at the same time giving a modern terminology to it which is still relevant for us today.

---

[7] The Klymene Principle, ch.II. - As far as Spinoza is concerned, I partly follow the outline of my argument given there.

But we should note the following: Spinoza does not only cumulate the various strings of the philosophy of his time. He also visualizes philosophy as something which is practically identical with ethics. For him, philosophy is a theory of the conditions according to which human life is being defined. And this definition can only be rational, if it is succeeding with respect to an ethical frame of references. Ethics itself, unfolds the conditions according to which the human striving can be realized. Humans are capable in this sense, of finding and conserving their own mode of being, if they act according to adequate knowledge. Hence, there is a close connection between freedom and insight. It is necessary therefore, to find adequate ways (inveniri) in an appropriate project which is to be designed by humans themselves. If Spinoza's „Ethics“ (1p34)[8] states that the power of God is his own *essence* (Dei potentia est ipsa ipsius essentia), then, for humans, one could add in an existential sense that the power of humans is their own *existence* (Humani potentia est ipsa ipsius existentia). And the adequate form of this existence is being prescribed in terms of the virtue which leads forward to blissful happiness. For humans therefore, virtue in this sense and power, are identical. (4d8)[9] Not only is this a mere re-formulation of stoic ideas, but Spinoza also tries to define substance (God = nature) as *causa immanens* of the world, but in terms of a twofold perspective taken according to whether the relationship between God and world is visualized under the substantial (real) perspective of God himself, or under the modal perspective of humans who represent a finite mode of what worldly exists. Obviously, the former perspective can be taken in speculative terms only, but because God represents himself completely in each of his attributes, it is possible that humans can grasp his (substance's) existence in principle, provided they have developed the adequate knowledge about this due to their adequate reflexion. In this sense, everything is in God (quicquid est, in Deo est), but the viceversa is also true.

On the other hand, reflexion itself is an outcome of the organization of substance: The constitution of the latter according to which it is productive with respect to the field of modi, is its organization in terms of attributes.[10] This means that God does not really produce attributes, but he (as substance) is *attributively organized* instead. Hence, God is *causa sui* only in so far as he produces everything what there is, but this is only true with respect to the finite perspective of humans. Ne-vertheless, a finite mode is in God, because it is a created mode within the totality of nature. But in this mode as its cause, it is only God who acts as immanent cause of permanency, it is not the totality of nature which is acting as this cause.[11] Hence, the central position of nature which is classified by Spinoza in a twofold way according to the classification given earlier by Averroes: He actually differs between *natura naturans* (the actively creating nature producing things) and *natura naturata* (the passively produced nature which is the outcome of processes performed by natura naturans). The former represents the productivity of God, the latter its result. *Nature is the form of mediation in which God acts upon the world as seen (and interpreted) under the modal perspective of humans*. But in reality, he does not think himself, because it is only the humans who do. (2a2) And therefore, he is not a spirit either.[12] So nature has an important role to play within the frame of references which constitutes the world. But as it is only humans that think, the basic concept of worldly orientation is *intersubjectivity* in first place, nature only in second. It is *reflexion*, and it is the *political form of communication* that determine adequate knowledge. But the latter can only be achieved, if the structure of the world in terms of its nature is uncovered and logically displayed. Hence, to study nature means to lay the ground for adequate knowledge, and in the end, for adequate *action* according to ethical principles. And this is a holistic viewpoint indeed.

---

[8] Throughout the text I use the English notation when referring to Spinoza's „Ethics“, the first number giving the part, the letter referring to the state of the proposition, in a somewhat self-explanatory manner, the second number following the usual listing of propositions.

[9] Although, for reasons of simplicity, we keep „power“ here for Spinoza's „potentia“, it should be noted that „potential“ (in the sense of capacity, ability, ...), in contrary to „potestas“ (power) is certainly a better translation. In so far I follow the opinion of the English translator of Negri's book on Spinoza: The Savage Anomaly, University of Minnesota Press, Minneapolis, Oxford, 1991 (= L'anomalia selvaggia: Saggio su potere e potenza in Baruch Spinoza, 1981). In Italian, there is a natural difference between the two expressed in the title (potere, potenza). The same is true for French (pouvoir, puissance). I claim this difference also for German (Macht, Vermögen). See the translator's remarks in Negri's book, op. cit., xi sqq.

[10] In W.Bartuschat: Spinozas Theorie des Menschen, Meiner, Hamburg, 1992, 66, a precise discussion of this aspect is given in more detail.

Spinoza follows in this sense the Cartesian approach of deducing physics from metaphysics.[13] But he is far more radical than Descartes. Gabriel Albiac has expressed this clearly: „.... nothing happens in Nature ... The conclusion imposes itself. It is only within the framework of a general theory of Nature that there will be room to understand all things (human or other) that it contains. (Hence) ... the *Ethics* ... is a physics, which is a metaphysics."[14] Spinoza himself mentions this aspect in a letter to Blijenbergh when he speaks of: „ ... Ethics, which, as everyone knows, must be founded on metaphysics and physics ..."[15] He seeks to derive his ethical theory from this basic understanding of nature in first place, the psychology of humans showing up within this framework as *human nature*, a particular case of nature in general. In this sense, every human action must be conceived as a manifestation of nature.[16]

Note that this is the celebrated aspect of determinism in Spinoza, though not as we know it: Instead of „determinism" it would be more adequate to speak of „necessitarianism", because nature is actually determining behaviour, but in an in-

---

trinsically contingent way. Hence, there is always a sufficient reason for everything which happens, and a kind of causal closure (consistency), but there is not necessarily a fixed „program" of processes. (We actually would argue today that physics is determining all processes in the world, but very much in terms of a basic framework, of some arena for processes. So having found out that in the end, „everything is physics", is certainly a true result and statement, but it is also completely unsatisfactory without having done the hard work - namely reconstructing the various levels of the hierarchy of complexity and following up the lines of mediation towards the phenomena actually being observed.)

This has a decisive consequence for *praxis* visualized as human existence: For Spinoza, the striving for conserving the own being (conatus in suo esse perseverandi), this characteristic kind of *eigen-being*, can be interpreted in terms of an ethics which unfolds the conditions under which this striving can actually be realized. Under them, therefore, the individual human being can indeed arrive at its own being (eigen-being).[17] Provided he/she applies what Spinoza calls the *true method* consisting of „the knowledge alone of the pure understanding, of its nature and of its laws. To acquire this method, you must first of all distinguish between intellect and imagination, or between true ideas and the others, that is fictitious, / false and doubtful ideas, and, absolutely speaking, all those that depend on memory alone."[18]

Hence, this is the kind of „ideal of a unified science"[19] Spinoza is aiming at. In principle, he seeks primarily to improve the character of human beings by improving their self-understanding. And he justifies this with the argument that it would bring humans the ultimate peace of mind as integral aspects of nature. Garrett notes that Spinoza, very much like Hobbes, conceives of human beings as „mechanisms in nature" that are motivated by self-preservation and individual advantage. They can improve their way of life by the mutual employment of reason. Hobbes however, tries to show how humans can optimize the chances for a good life by instituting political constraints on their passions. Spinoza's aim is far more ambitious: He sets out to show how humans can achieve a way of life that largely transcends the mere transitory desires leading to an autonomous control over passions.[20]

It is in this way that Spinoza's monistic and naturalistic system speaks „most cogently and persuasively to the twentieth century."[21] In fact, it is doing so by explicating three basic aspects which have become very important for modern research in recent times: visualizing a transcendental materialism with physics „at the bottom", in a radical approach to interdisciplinarity, and aiming at mediations leading up to ethics and politics.

---

[17] Bartuschat, op.cit., x. (par.) - See also for an unfolding of intersubjectivity based on this ethics: 186 (referring to 4p18) and 200. With strong Sartrean connotations actually, because such a human can be easily visualized as one „who is able to become what he/she actually is".
[18] As quoted by Gabbey, op.cit., 171sq. - referring to the epistle (37) to Bouwmeester.
[19] Curley, op.cit. (title of ch.1 sect. 2), 4, 6.
[20] Garrett, op.cit., 267sq., par.
[21] D. Garrett in his introduction to the Cambridge Companion, op.cit., 2.

As far as physics is concerned Spinoza writes to Tschirnhaus:

> By physics I understand nothing else than the science of the universe demonstrated a priori by the rigorous method of the mathematicians and confirmed a posteriori by the most evident experiences which even convince the imagination ... This science is truly divine. One here exposes the laws ... according to which everything produces invariably its effects. The knowledge of this science liberates us also of innumerable prejudices ... In this way, through the mediation of the true physics, one becomes so to say a completely new man and one is regenerated philosophically. ... Ultimately *thanks to physics* we are prepared for still more important knowledge.[22]

Hence, the program of the „Ethics" is one which confirms our view today as we have introduced it in the beginning: the first point being to know our nature as good as possible, desiring to perfect it, and the second to know as much of the nature of things as is necessary.[23]

However, the radical kind of interdisciplinary approach is also very much on our modern line: Spinoza's necessitarianism (being based on a firm view of ontological monism) leads him to abolish the classical divisions (of *prudentiae* and *scientiae*) „making the principal subjects of the Ethics and the two political Treatises formal impossibilities within a Peripatetic perspective."[24] This is what Negri actually means when speaking of Spinoza as „anomaly": „Spinoza is the anomaly. The fact that Spinoza, atheist and damned, does not end up behind bars or burned at the stake, ... can only mean that his metaphysics effectively represents the pole of an antagonistic relationship of force that is already solidly established: The development of productive forces and relations of production in seventeenth-century Holland already comprehends the tendency toward an antagonistic future."[25] In his striving for an *experientia sive praxis*[26], Spinoza aims at a theory of the world, beginning with physics, ending up with social systems, proposing educational means of eventually approaching a state which, in an equilibrium of the individual and the institution, harmonizes human spirit in order to let it participate in the larger harmony which is expressed in the material attribute of godly substance. This is obviously the reason for the recent reception of Spinoza's in explicitly Marxist terms (notably in Althusser). It is Tosel who recognizes an *ethico-political* consequence emerging from Spinoza's approach, basing it on definite materialistic *philosophemes* and a collection of motives generic for Marxist theory: „L'onto-théologie est éthico-politique: Dieu, c'est-à-dire la légalité d'une Nature immanente, se traduit humainement dans l'immanence d'une société ratio-

---

nellement réglée d'hommes capables de penser et agir. Et l'éthico-politique est à son tour ontologique: l'homme libre est une possibilité de la nature anonyme."[27] He lists then a number of appropriate philosophemes: Nothing comes from noth-ing(ness) - *ex nihilo nihil fit* -, everything has a reason which can be illuminated by its foundation in terms of possibility, the real has objectivity which can be il-luminated by science, hence, philosophy is also scientific and deals with the fun-damental structures of what there is, dealing also with itself, in order to produce knowledge about itself and about human thinking, hence philosophy-science de-termines itself as a science of humans, being based on the fundamental thesis that humans are themselves nature, not an empire within an empire, this science of humans is scientific ethics, that is re-organization and re-orientation of human life.[28] In this sense, Tosel shows that for Spinoza, all of reality emerges from another reality which is basically material, the unity of which is immanent in the variety, and coincides with its own space of productivity. Hence, cosmology, physics, and logic are the basic fields for constituting anthropology.[29] The results following immediately from this are exactly the topics which are our interest to-day when trying to find a new synthesis of philosophy, science, arts, and praxis (of which the first three are a part). And we can realize the strong existentialistic connotation being handed down to us from Spinoza: Because of the elements just mentioned, philosophy has a fundamental interest in liberating humans, in aiming to a positive liberty of individuals against all kinds of heteronomy. Philosophy is theory and praxis of autonomy. And in being a science of humans, philosophy is also an instrument of critique, especially of critisizing all authorities which do not base their principles on communicative arguments. Research about life is therefo-re research of (and about) *free* life in the first place, of a life which has no other generic principle than its physico-psychic unity. For this, nature is the original reality, and as such nature is intelligible, and it is anterior to all thinking.[30]

## 3  The Principles of Substance

In order to clarify the concept of substance, it might be the best to look for its re-lation to the fundamental categories of space and time in more detail: Essentially, the attribute of extension can be visualized as space itself.[31] The actual difference between Leibniz and Spinoza can be phrased then in terms of the relationship between objects and space. The former explains away the region and stays with the relations among bodies, taking the region as an alternative way of expressing facts about them. The latter explains away the objects and stays with the region. In referring to a famous example, Bennett formulates: „If there is (...) a pebble in region R, what makes this true is the fact that R is *pebbly* (which) stands for a certain monadic property that a spatial region / can have. If the pebble moves (...), what makes this true is the fact that there is a continuous change in which regions are pebbly: The so-called movement of a pebble through space is like the so-called movement of a panic through a crowd."[32] Note that this gives a relatively clear explanation of how we should visualize *motion-in-itself*. In this sense, time does not remain as a fundamental category. Spinoza states in his 12[th] letter that „ ... *tempus* (is) nothing but a mode of the imagination which ought to mean that in a true fundamental account of the whole of reality th(is) concept ... would not be used. (Hence,) all our measures - of time and space and of things spatial and temporal - are superficial and „imaginative", and not part of the basic, objective story."[33]

---

[27] A.Tosel: Du Matérialisme de Spinoza, Kimé, Paris, 1994, 18.
[28] Ibid., 132 par.
[29] Ibid., 133, 135, par.
[30] Ibid. 187sq., 190, par.
[31] See e.g. J. Bennett: Spinoza's Metaphysics, in: D.Garrett (ed.), The Cambridge Companion to Spinoza, op.cit., 61-88, here: 69.

Although this concept is widely agreed upon, at the same time, it has been the target of many discussions. It can be shown however that Spinoza ties time to the imagination (2p44s), basically referring it to human perceptions of varying motions of bodies.[34] Hence, temporality actually emerges in the transition from infinity to finite modes (of thinking) as laid down in 2p8 and its corollary.[35] In fact, as Bartuschat mentions explicitly in his discussion at this point, the corollary has the important function of marking the transition to finite modes, and it is here where Spinoza actually changes the perspective of description in his „Ethics".[36]

Obviously, if temporality is part of the human world view in modal terms, then there is no problem of determinism. But still, the question is as to the actual structure of the human mode of being which is also the human mode of perception. Of course, Spinoza's identity theorem (2p7) is the main topic around which this question centres. If „thought is co-extensive with materiality"[37], then the ethical implications of Spinoza rest basically on the fact that human thinking is not a mere passive reflexion of the body's vicissitudes: „As well as being mapped onto a segment of the material world, the mind is inserted into the totality of thought."[38] Hence, a singular individual is not a self which also strives for some-thing, but a singular individual is a self-in-striving, and it has a *project structure* (very much in the sense of modern existential philosophy). But this project which in the end shows up as the dynamical core of existence, is a representation of the human's finite mode of being, which points toward the adequate form of knowledge humans can have, because it is this knowledge in which they actually conserve their own being. And the generic form of realizing existence is *acting* according to the results of adequate reflexion.[39]

Hence, human reflexion means nothing else but explicating the symbolic traces in existence which intrinsically point to the true mediation of what there is with substance, visualized in terms of the latter's attributes. Modelling the world in human terms means modelling the configuration of (worldly) attributes. Obviously, the latter cannot be modelled without modelling its foundation (substance) at the same time. This is the point where *sceptical* philosophy is clearly based and dependent on *speculative* philosophy. In fact, human reflexion has the actual task of organizing the infinite[40] by means of the finite. And both of them are „knotted to each other" in terms of the consequences of the identity theorem 2p7: „Mental items can be mapped onto bodily items in a way that preserves causal connectedness."[41] Hence, if M is the category of mental states, and B is the category of bodily states, respectively, then there is a functor $\varphi$: B $\to$ M (preserving identities and compositions). If so, probably, we could expect that the diagram

$$M_1 \to M_2$$
$$\uparrow \qquad \uparrow$$
$$B_1 \to B_2$$

commutes, for any pair of bodily states, and mental states, respectively, when the horizontally parallel arrows $\to$ refer to morphisms in the respective categories, and the vertically parallel arrows $\uparrow$ define generic „liftings". (In fact, one would expect appropriate isomorphies to hold here.)[42] These relationships secure that humans are capable at all to adequately interpret the symbolic traces intrinsic to the world they observe. And this is the reason why speculative philosophy is not only a necessary action to be undertaken, but also one which might result in sufficient insight.

---

[39] Bartuschat, op.cit., xi, 132sq.

[40] Negri, op.cit., 52.

[41] This is an alternative formulation by Bennett, op.cit., 78.

[42] That is we would expect that there are generic projections p and q such that the compositions with the respective liftings, a and b say, reproduce the various identities: p o a = 1 (M), a o p = 1 (B), and the appropriate for q and b. - The version of the theorem which Bennett gives is not very clear when starting with the idea of parallelism from the outset (as he actually does). He says that mode identity would mean that „if M is correlated with B under parallelism, then M = B." (Ibid., 79) In fact, „identity" in the theorem 2p7 refers to the „order and connection" of ideas on the one hand, and of things on the other. („Ordo, & connexio idearum idem est, ac ordo, & connexio rerum.") It does not mean that ideas and things are actually the same. It is true however that „the role of the attributes is to combine with the transattribute modes to get the latter into a form in which we can think them." Indeed, the „attributes let the / modes come through. It is as though the modes were words written in a script to which intellect is blind, and the attributes make the message of the modes accessible to intellect by reading them aloud, expressing them." (Ibid. 87sq.) This relates to the aspect of self-narration which I have discussed at another place. See my „Prosperos Buch oder Echolot der Materie. Zum hypothetischen Natursubjekt bei Ernst Bloch: Bilanz & Ausblick." (VorSchein, Nürnberg, 15, 1997, 40-57.) Reflecting about this there is the possibility of discussing it in terms of relating modes to (mathematical) logoi - in the sense of René Thom -, and attributes to (mathematical) topoi, respectively. I have discussed this in more detail

As to the semiological re-construction of these symbolic traces, Deleuze[43] has given a classification of signs which are relevant for the understanding of this mediative structure of the world according to Spinoza. He introduces four principle types of scalar signs, and two of vectorial signs, in the following manner: *Scalar* signs are of *indicative* type according to whether they deal with sensory or perceptive physical effects, indicating human nature rather than anything else, of *abstractive* type according to whether our nature, being finite, retains some selected characters from what affects it, of *imperative* type according to whether effects are taken for ends, or ideas of effects for causes, actually producing moral effects, and of *hermeneutic* type, according to whether humans imagine suprasensible beings as an enlarged image of what effects them. Hence, these types of signs define sensible indices, logical icons, moral symbols, and metaphysical idols, respectively. *Vectorial* signs can be identified with affects and are basically of an *augmentative* type, if they carry positive connotations, or of a *diminuitive* type, if they carry negative connotations. The common characteristics of all these signs are associability, variability, and equivocality or analogy. The vectorial signs may be combined in their effect in order to define explicit gradients of social fields.[44]

Deleuze points out that „(if) signs, like words, are conventional, it is precisely because they act on natural signs and simply classify their variability and equivocity: conventional signs are abstractions that fix a relative constant for variable chains of association. Signs do not have objects as their direct referent. They are states of bodies (affections) and variations of power (affects), each of which refer to the other. Signs refer to signs."[45] Thus signs are also effects, and effects refer to effects as signs to signs. They are actually „consequences separated from their premises." Deleuze uses here an optical metaphor: „Effects or signs are *shadows* that play on the surface of bodies. The shadow is always on the border. It is always a body that casts a shadow on another. ... Signs are *effects of light* in a space filled with things colliding into each other at random."[46]

In Spinoza, the determining oppositions to signs are common notions or concepts. They are primarily based on structures, and modes are geometric but fluid structures that are permanently being transformed. Hence, structure is rhythm, that is, the linking of figures that compose and de-compose their relations. In this sense, modes can be visualized as projections of light: „Or rather, the variations of an object are projections that *envelop* a relation of movement and rest as their invariant (involution)."[47] So they are also *colours* or colouring causes: „Colors enter into relations of complementarity and contrast, which means that each of them, at the limit, re-constitutes the whole, and that they all merge together in whiteness (infinite mode) following an order of composition, or stand out from it in the order of de-composition."[48] Hence, the whole „Ethics" can be read in terms of the scholia which constitute a book of signs „which never ceases to accompany the more visible *Ethics*, the book of the concept ..."[49] Signs and affects, or concepts, shall be transgressed in the end, by arriving at essences or singularities, percepts. No longer signs of shadow or of light as colour is important then, but light *in itself and for itself*.[50] Hence, the adequate form of reflexion required here is simply a hermeneutic approach to nature. For eventually achieving this, the actual transition from logic to hermeneutic must be based on the space of free play which is intrinsic to the mediation of world and substance, modal and real existence, as it is characterized by the parallelism which can be represented in mathematical terms. This is an aspect which practically unifies the ontological and epistemological domains.

For Spinoza therefore, substance is what is in-itself and what can only be comprehended by itself and out of itself. This means that it is something whose concept does not need the concept of anything else in order to be formed. Hence, substance is its own reason (causa sui), and its essence involves its own existence. But humans, being a finite mode of this infinite substance, can only perceive their world in terms of attributes of this substance, but not substance itself. In fact, their world is actually being defined in terms of these attributes of which there are infinitely many, of which humans however can only perceive the two which fall into their mode of being: matter (res extensa) and mind (res cogitans). Note that not only is the world not the same as substance (hence all that there is for humans is only one special aspect of all that there is in reality), but humans are also permanently modelling the attributes, in a kind of recursive approximation, rather than perceiving them as they actually are. Hence, the relationship between the real perspective of substance and the modal perspective of humans is difficult to visualize: E.g., although the infinite substance is undivisable (has no parts), humans actually perceive parts of attributes, but this difference is only given in modal, not in real terms. (1p15s)

Hence, for humans, substance is the foundation of being, and as such it is *nonbeing*. It is not nothing, because the world, as visualized in modal terms, has eventually emerged from substance. In a way, we can say that the world has been produced by substance out of a field of possibilities. Hence, worldly objects come into existence by some initial emergence of the world which is thus an exterio-

---

[47] Ibid., 24sq. (par.)
[48] Ibid., 25.
[49] Ibid., 28.
[50] Ibid., 30 (par.)

rization of substance (in the sense that substance unfolds its organizational struc-
ture - which is not really a process when visualized under the real perspective of
substance, but rather an equivalent self-representation of substance itself). Note
that when we as humans, are talking about this emergence, the very language we
have to use points to a dynamical process of some kind. The reason for this is that
we have to think according to our usual (modal) terms of reflexion which involve
temporality one way or the other. But, in reality, substance within this picture, is
not actually triggering dynamical processes, it is already structured in a way that
it is differentiating itself in terms of infinitely many self-representations, of which
one is what we interpret as our world. Hence, motion in terms of substance is ill-
defined. It is rather that substance is constituted as potentially self-moving in the
sense that it is in a state of permanent self-fluctuation which represents an intrin-
sic sort of motion, an abstract motion (a motion in-itself), and a potential for con-
crete, worldly motion, at the same time. It is in this way that the intrinsically dy-
namical constitution of substance points to a concept of freedom which basically
means „freedom to eventually produce a structured world". This could be inter-
preted as a kind of absolute freedom which in itself can be taken as a practical
definition of substance.[51]

The world is therefore a kind of deficient state of substance as it appears only in
terms of restrictions (as being primarily finite). Substance however, as foundation
of being, is itself without foundation. Hence, it is constituted in terms of self-
reference, and it propagates aspects of this self-reference into the world. In this
sense, substance is beyond space and time, it is basically non-local, it may be vi-
sualized as „pre-geometry". At the same time, the world as the product of sub-
stance is constituted in a transcendental sense, because there is an immanent ten-
dency of the worldly towards returning to its own origin. (Under the real per-
spective of substance this means that substance has itself the tendency to re-
integrate its own unfolding (representation) into itself, which is in fact nothing but
an alternative expression of its own totality. In principle then, the world can be
re-interiorized again, and it is this final stage of development, in which worldly
existence is sublated again in its original (primordial) unity - very much in the
threefold Hegelian sense.

# 4 Pre-Geometry and TOE

Since the advent of general relativity theory, the strive for a unification of science
(philosophically speaking: for grasping the foundations of physics and hence the
foundation of the world) has entered a new phase of intensity. For Einstein him-self, it appeared appropriate to look for a unification in terms of gravitation and electromagnetism, the two fundamental forces (or interactions) at the time. To be more precise: Einstein's idea was to actually include electromagnetic fields in his gravitational field equations by introducing them (by means of their potentials) into the metric components of space-time, in first place. As Wolfram Voelcker, Can Yurtoeven, and myself have shown at another place[52], this can be interpreted in terms of *visualizing substance as space-time geometry* in the sense of general relativity, representing gravitation, at the same time. (Indeed, as Lewis Feuer has pointed out earlier[53], it is quite certain that Einstein developed his theory under the impression of discussing Spinoza's philosophy when in Prague.) The subsequent advent of quantum theory however, to which Einstein himself contributed some fundamental insight, led to a much more complex situation, because not only did two more forms of interaction (weak and strong fields) turn up, but the basic interpretations of relativity on the one side and quantum theory on the other, differed according to their respective (worldly) domains: on the one hand, relativity being defined in terms of a four-dimensional space-time manifold with a Lorentz metric (of signature -2), quantum theory on the other, showing up as being defined in terms of a high-dimensional Hilbert manifold with a positive-definite (Euklidean) metric. Hence, one ended up with two disjoint domains claiming to explain the same universe. So, although in the meantime various intermediate successes have been celebrated (such as the unification of the electroweak force, or the approach towards a *grand unification* of the latter with the strong forces: GUT), the ultimate goal, the unification of all forces (and matter) with gravitation in a *theory of everything* (TOE) has not yet been achieved.

During the last three decades, a basically different approach to unification has been put forward going back to an old idea of John Wheeler's: If it is not possible to unify the competitors within the world, it might be possible to unify them *outside* the world, the basic idea being to introduce an abstract mathematical structure from which space and time (and matter) as fundamental categories of the world could be eventually *derived*. It is quite straightforward in fact, to notice the connotation of substance here: If we define our world in terms of fundamental categories such as space and time (and matter), then everything outside the world from which we might be able to derive these fundamental categories is the *foundation* of the world and as such it is *non-being*. Hence, the idea is to visualize the world as a variety which has become out of a primordial (actually pre-worldly) unity. If so, then the next step, namely to formulate this the other way round: that the world *is* in fact this primordial unity as being observed as a becoming variety by its members who have restricted means of perception, is relatively small. Contrary to what Einstein thought, space-time-matter would not be substance it-

---

self, but only the latter's worldly attribute. And what is „before" (and external to) the world, *pre-geometry*, would gain the connotation of a substance.

The question is how to reasonably approach such a conception in more detail. Because, technically, this would point to achieving a unified context for both gravitation and quantum theory anyway. A useful motivation for „quantum gravity" of that sort has been given by John Baez recently[54]: There are three fundamental length scales, defined in terms of three physical constants coming from both regions in question ($\hbar$, the Planck constant divided by $2\pi$, Newton's gravitational constant G, and the velocity of light c, respectively), which are important for relativity and quantum theory, at the same time. One is the *Planck length* $l_P = (\hbar G/c^3)^{1/2}$, of roughly the order of magnitude of $10^{-35}$ m. For length scales smaller than this, quantum gravity would be the appropriate tool to describe the physics there. Another one is the *Compton wavelength*, $l_C = \hbar/mc$, which basically indicates that the measuring of the position of a particle of mass m precisely within one Compton length, requires energy which is able to create another particle of the same mass. Hence, this length scale is characteristic for effects in quantum field theory. Finally, there is the *Schwarzschild radius* which basically defines the horizon of a black hole which has been formed by a collpasing star of mass m: $l_S = Gm/c^2$. (All constants up to numerical factors.) If m is now the Planck mass itself ($m_P = (\hbar c/G)^{1/2}$), then $l_C = l_S = l_P$. Hence, we would suppose that at the Planck scale, both domains of physics should show up in a somewhat unified way.

Thinking of the traditional division of classical relativity, and quantum theory, respectively, it is quite natural to ask whether the continuum picture of space-time is only an approximation which inevitably breaks down when approaching the Planck scale. And if so, whether there are constituents of space and time which show up according to some scheme of quantization, such as to construct quantum operators with discrete eigen-values. The microscopic structure of space and time would be determined then by eigen-values and eigen-vectors of purely geometric operators, and the macroscopic superposition of these would show up as the well-known space-time continuum (as a limit for large values). But this would also mean to abandon any underlying space-time structure we have got accustomed with (both in relativity *and in quantum theory* which is also based on a classical space-time background). As Ashtekar and Krasnov have pointed out, „ ... to probe the nature of quantum geometry, one should not begin by *assuming* the validity of the continuum picture; the quantum theory itself has to tell us, if the picture is adequate ..."[55]

For their approach, referred to as „loop quantum gravity", it is the goal therefore, to find a background-free quantum theory with local degrees of freedom propagating causally. This is also true for related approaches (e.g. topological quantum field theory (TQFT) etc.), but it is *not* for the celebrated *theory of superstrings*. The latter has the same objective though, in general terms, namely looking for a *pre-geometric TOE* of some kind. It is however, a chiefly *perturbative* theory meaning that it starts from a given space-time background (very much in the tradition of quantum theory), performing a perturbation expansion as power series expansions of the string coupling constants similar to those used in quantum field theory, and extracting perturbational modes corresponding to physical particles (or particle fields). To that end, all superstring theories contain a massless scalar field called the *dilaton* that belongs to the same supersymmetry multiplet as the graviton. It determines a suitable string coupling constant on which the perturbation expansions can be based.[56] However, the big problem is with the background (whose existence contradicts the principle of diffeomorphism invariance which is one of the central results of general relativity). As Carlo Rovelli has put it, keeping a background means to describe the motion of physical entities on a non-dynamical stage instead of describing the dynamics of the stage itself.[57]

There are also other problems of which the dimensionality of the appropriate superstring spaces might be the most relevant. Although the Kaluza-Klein tradition in physics has its merits (and actually impresses by its beauty), and although the problem of ending up with five ten-dimensional theories rather than with one fundamental theory seems to be resolved now by introducing certain dualities among them, the idea being that there is some large moduli space of consistent vacua of a single underlying theory (called M theory by now) containing also a theory of eleven dimensions, it is nevertheless very unsatisfactory - to say the least - that a mechanism is to be (practically arbitrarily) postulated in order to explain why the dimensions should curl up to microscopic spaces (which is actually a flagrant breach of the rule as given in terms of the theorem of sufficient reason) so as to leave us with the four dimensions as we know them. In fact, as we have seen already, this could be accounted for in a much more straightforward way, thinking of our world as an attribute of substance (or: alternatively: as a visualization of the world's foundation). Finally: leaving aside the derivation of the black hole entropy with the exact Bekenstein-Hawking coefficient, which is certainly a success in its own right, also the *experimental evidence* according to the expectations determined by the standard model, is not very well supported at all, because, as Rovelli mentions in his review paper, most testable predictions have all failed to be confirmed so far. Rovelli argues that comparing this lack of success with the considerable success the standard model had in the past, leads to the assumption that there might be some principal limit to further following this path.[58] Things have not been improved since then, as the recent Minnowbrook symposi-

---

on has shown.[59] This is the reason why within this paper we will concentrate on the other approach (loop quantum gravity and related) rather than on superstrings or M theory. But it should be noted that in principle, the latter approach is also on the line with the concept of substance, and so it is worth being mentioned here.

The approach put forward by Ashtekar and others is mainly based on introducing a new kind of connection which replaces the Levi-Civita connection of general relativity. The idea is to simplify procedures and to bring out more clearly the kinematical similarity of this ansatz with Yang-Mills theory. (Recall that according to the standard theory, Yang-Mills fields show up as gluons in their condensed state being responsible for the quark confinement. In this model then, matter is being visualized as consisting of quarks and leptons together with their Yang-Mills components, namely showing up as 36 quarks (coming in six flavours and three colours - times two for their anti-particles), six leptons (caring primarily for weak interactions), and twelve Yang-Mills components (eight for the gluons plus four for electroweak interactions). These families come typically in three generations. Usually, this can be alternatively described in terms of group symmetries, for the GUT of the characteristic form U(1) x SU(2) x SU(3), the idea being that the subsequent breaking of symmetries would have actually produced the interactions within the universe as we know them. In particular, within string theory, symmetries may be visualized as side-effects of (string) oscillations in hyperspace, and it is tried to combine all these components as metric components on this very hyperspace - which is nothing but the original idea of Kaluza, only applied to higher dimensions.[60])

Coming back to loop quantum gravity: Starting with the 3+1 split of the metric, the phase space of general relativity can be described in terms of a three-dimensional manifold M which is compact and without boundary, and a smooth real SU(2) connection $A^i_a(x)$, as well as a vector density $E^a_i(x)$. Here x refers to co-ordinates on M, the letters a,b, ... = 1,2,3 to spatial, the letters i,j, ... = 1,2,3 to internal indices. The relation to the conventional symbols is secured by the equations

$$g\, g^{ab} = E^a_i\, E^b_i,$$

where g = det $g_{ab}$, and

$$A^i_a(x) = \Gamma^i_a(x) + \gamma\, k^i_a(x),$$

$\Gamma$ being here the spin connection associated to the local triad for whose labelling the internal indices can be visualized as being used, k referring to the extrinsic curvature of the constant time 3-surface.[61]

Note that this space-time-split establishes already a theoretical choice which is very relevant for the philosophical implications as to the significance of time: What changes in general relativity in dynamical terms is not the 4-distances within space-time, but rather the 3-distances within spaces as being nested in space-time. Hence, the dynamics is essentially one of 3-dimensional Riemannian spaces. This idea going back to John Wheeler in the sixties is discussed in Julian Barbour's new book[62], where Barbour points out that the „key geometric property of space-times that satisfy Einstein's equations reflects an underlying principle of best matching built into the foundations of the theory."[63] The time separation of spatial slices shows up here as what Barbour calls a *distinguished simplifier*, as an ordering principle for making unfoldings simple.[64] If time is being visualized as a mere ordering principle, then, in philosophical terms, we are left with space as an attribute. Note however, that the *dimensionality* of space is only a finite representation then, which does not reflect the true nature of space, but only our modal attitude towards it with a view to spatial ordering.

The Ashtekar ansatz is also invariant under local SU(2) gauge transformations, three-dimensional diffeomorphisms of the manifold on which the fields are defined, as well as under (coordinate) time translations generated by the Hamiltonian constraint. The full dynamical content of general relativity is also captured by the three constraints that generate these gauge invariances.[65] So what we can do now is to compare the configuration variable of general relativity as known from gauge theories with the SU(2) connection A on a spatial 3-manifold, and the canonically conjugate momentum E with the Yang-Mills „electric" field. Physically, the latter is essentially the triad and carries all information about space. This is where in quantum theory, the gauge invariant *Wilson loop functionals* are coming in: They are the path-ordered exponentials of the connections around closed loops. Hence, the name for the theory (loop quantum gravity). We will come to that in a moment.

---

[61] Note that the Einstein summation convention does not apply to internal indices. The parameter $\gamma$, called the Immirzi parameter, is usually chosen to be equal to the imaginary unit for recovering the Ashtekar standard notation. I am following here the presentation in C.Rovelli: Loop Quantum Gravity, Living Reviews in Relativity, www.livingreviews.org/Articles/Volume1/1998-1rovelli, Max-Planck-Gesellschaft (Potsdam), 18sqq. See also: Ashtekar, Krasnov, op.cit.

[62] J.Barbour: The End of Time (The Next Revolution in Our Understanding of the Universe), Weidenfeld & Nicolson, London, 1999. - Note that for the Einstein vacuum equations $R_{ab} = 0$, with space-time as $M = R \times S$, where S is the $(t = 0)$-slice of M, $R_{0b} = 0$ are the constraints on the initial data, and the remaining equations give the evolution in time. The physical states are given as the subspace of diffeomorphism invariant states that are annihilated by the constraint corresponding to $R_{00}$. The equations expressing this fact are the Wheeler-deWitt equations, which, in the book of Barbour's, take a central position therefore.

[63] Ibid., 176.

[64] Ibid., 180.

[65] Rovelli, op.cit. (n.61), 19.

Note however another relationship first: Usually, the kinematics of quantum theory is defined by an algebra of operators on a Hilbert space. The outcome of the physical theory will depend on the connection which can be uncovered between operators and classical variables, and on the interpretation of the Hilbert space as a space of quantum states. The dynamics is determined by a Hamiltonian (in general relativity called quantum constraint) constructed from the operators. The idea is now to express quantum states in terms of the amplitude of the connection, given by some functional of the type $\Psi(A)$ in the sense of Schrödinger. Functionals of this kind form a space which can be made into a Hilbert space provided a suitable inner product is being defined. Having done this, we can express the amplitude as $\Psi(A) = <A \mid \Psi>$ in Dirac notation. Is now $H$ our Hilbert space, then we define $H_0$ as its subspace formed by states invariant under SU(2) gauge transformations. Then it can be shown that an orthonormal basis in $H_0$ is actually a *spin network* basis.

What is a spin network? Basically, spin networks are graphs with three-valued nodes and spin values on the edges, their states being denoted by $\mid \Gamma>$. To take the norm refer to the mirror image of a graph and tie up the ends, forming a closed spin network $\Gamma\#\Gamma^*$ of value $V$ such that

$$<\Gamma \mid \Gamma> = V(\Gamma\#\Gamma^*)$$

with

$$V(...) = \Pi \ (1/j!) \ \Sigma \ \varepsilon(-2)^N,$$

here j being the edge label, N the number of closed loops, and $\varepsilon$ referring to the intertwining operation taking care of permutations of signs. The product is to be taken over all edges, the sum over all routings. Hence, the networks can be visualized in diagrammatic form such as to represent the underlying „spin dynamics". For instance, a diagram with vertices a, b, c, and i, j, k within the region of intertwiners such that i + j = a, i + k = b, j + k = c, can be interpreted as two particles with spin a and b which produce (create) a particle with spin c. Spin interactions of this kind can lead to the creation of new structures: Take a large part of the network (effectively representing a part of the universe) and detach from this small m-units, and n-units, respectively, as „free ends". The outcome of their tying up to form a new structure can be estimated in terms of a probability for the latter having spin number P, say. This turns out as being basically the quotient of the norm of the closed network and the norm of the network with free ends (times some intertwining operations). The *spin geometry theorem* tells us then that when repeating this procedure and getting the same outcome, then the new quotient is proportional to (1/2) cos $\theta$, where the angle is one which is taken between the axes of the large units. Hence, it is possible to show that angles obtained in this way satisfy the well-known laws of Euklidean geometry. Or, in other words: This purely combinatorial procedure can be used to actually approximate space from a *pre-spatial* structure which is more basic. This idea has been due to Penrose who originally tried to base his concept of *twistors* on spin networks. In looking for primary concepts of an abstract structure from which space (and eventually space-time) could be approximated, starting from purely combinatorial elements, he began with looking more closely to the implications of angular momentum as to the re-construction of space.[66] The consequences of this approach have remained relevant until today, although twistors are not in fashion nowadays.[67]

A generalization of spin networks and a connection with knot theory has been achieved more recently by Carlo Rovelli and Lee Smolin referring to their concept of quantum gravity: They start with loops from the outset and show that since spin network states <S | span the loop state space, it follows that any ket state |ψ> is uniquely determined by the values of the S-functionals on it, namely of the form

$$\psi(S) := <S \mid \psi>.$$

To be more precise, Rovelli and Smolin take embedded spin networks rather than the usual spin networks, i.e. they take the latter plus an immersion of its graph into a 3-manifold. Considering then, the equivalence classes of embedded oriented spin networks under diffeomorphisms, it can be shown that they are to be identified by the knotting properties of the embedded graph forming the network and by its colouring (which is the labelling of its links with positive integers referring to spin numbers).[68]

When generalizing this concept even further, a network design may be introduced as a conceptual approach towards pre-geometry based on the elementary concept of *distinctions*, as Louis Kauffman has shown.[69] In particular, space-time can be visualized as being produced directly from the operator algebra of a distinction. If thinking of distinctions in terms of 1-0 (or yes-no) decisions, we have a direct link here to information theory, which has been discussed recently again with a view to holography.[70] Ashtekar and Krasnov have noted this already when deriving the

---

[71] Hence, the reference to Wheeler's „It from Bit". (Paola Zizzi has tried to generalize this within the conception of inflationary cosmology, and terms this „It from Qubit".[72])

Generalized spin networks can be used now for lattice gauge theory and for non-perturbative quantum gravity. In the former case, they turn out as products of Wilson loops. In the latter case, it can be found that the space of diffeomorphism invariant states is spanned by a basis which is in one-to-one correspondence with embeddings of spin networks.

Note that according to the standard terminology, a loop in some space $\Sigma$, say, is a continuous map $\gamma$ from the unit interval into $\Sigma$ such that $\gamma(0) = \gamma(1)$. The set of all such maps will be denoted by $\Omega\Sigma$, the loop space of $\Sigma$. Given a loop element $\gamma$, and a space x [73] of connections, we can define a complex function on xx $\Omega\Sigma$, the socalled *Wilson loop* such that

$$T_A(\gamma) := (1/N) \, Tr_R \, P \exp \int_\gamma A.$$

Here, the path-ordered exponential of the connection $A \in x$, along the loop $\gamma$, is also known as the holonomy of A along $\gamma$. The holonomy measures the change undergone by an internal vector when parallel transported along $\gamma$. The trace is taken in the representation R of G (which is the Lie group of Yang-Mills theory), N being the dimensionality of this representation. The quantity measures therefore the curvature (or field strength) in a gauge-invariant way.[74]

In Topological Quantum Field Theory and in Conformal Field Theory, the path integral definitions are mainly based on (quantum) Chern-Simons theory. (Recall that if A is a connection 1-form for a gauge group, then the quantum Chern-Simons theory has the path integral

$$Z = \int d\mu(A) \exp (i\hbar/4\pi \, S^{cs}(A)),$$

where $S = \int Tr \, (A \wedge dA + 2/3 \, A \wedge A \wedge A)$ is the Chern-Simons action on a compact 3-manifold $\Sigma$.) Over a given loop $\gamma$, the expectation value $< T(\gamma)>$ turns out to be equal to a knot invariant (the „Kauffman bracket") such that when applied to spin networks, the latter shows up as a deformation of Penrose's value $V(\Gamma)$. This is mainly due to the fact that

$$< T(\gamma)> = K^k(\gamma) = (1/Z) \int d\mu(A) \exp (...) \, T(\gamma, A).$$

So, for any spin network $\Gamma$ (replace $\gamma$ by $\Gamma$), the old relation holds up to regularization. Hence, spin networks are deformed into quantum spin networks (which are essentially given by a family of deformations of the original networks of Penrose labelled by a deformation parameter $q = \exp (4\pi/(k+2))$ for the Chern-Simons case). The latter may be understood as built up from the representation theory of quantum groups which are deformations of Lie algebras (namely Hopf algebras).[75]

Similar results can also be shown in terms of a categorial approach introduced by Segal and Atiyah.[76] Note also that the CS invariant is important when having a non-zero cosmological constant $\Lambda$, because there is an exact physical state of quantum gravity given by $\Psi_{cs}(A) = \exp (k/4\pi \, S^{cs}(A))$, where k is actually related to Newton's constant by $G^2\Lambda = 6\pi/k$. This state can be shown to reproduce $K^k(\Gamma)$ above.[77]

There is also a simplicial aspect to this: Loop quantum gravity provides for a quantization of geometric entities such as area and volume. The main sequence of the spectrum of area e.g., shows up as $A = 8\pi\gamma\hbar G \sum_i (j_i(j_i + 1))^{1/2}$, where the j's are half-integers labelling the eigenvalues. (Compare this with the remarks on black holes above.) This quantization shows that the states of the spin network basis are eigenvalues of some area and volume operators. We can say that a spin network carries quanta of area along its links, and quanta of volume at its nodes. A quantum space-time can be decomposed therefore, in a basis of states visualized as made up by quanta of volume which in turn are separated by quanta of area (at the intersections and on the links, respectively). Hence, we can visualize a spin network as sitting on the dual of a cellular de-composition of physical space.[78]

As far as the dynamics of spin networks is concerned, there is still another, more recent approach, which appears to be promising as to the further development of topological aspects of quantum gravity (referred to as TQFT). In setting out to develop this new ansatz[79], John Baez notes that there are basically only two new ideas involved in loop quantum gravity. One is the insistence on a background-free approach. The other is to base the theory on the aspect of parallel transport rather than on the metric. Spin networks are at the basis of this approach. But although originally, Penrose thought of them in terms of describing the geometry of space-time, they really turn out to describe the geometry space much better. The idea of Baez is therefore, to supplement loop quantum gravity with an appropriate path-integral formalism. While in traditional quantum field theory, path integrals are calculated using Feynman diagrams, he would like to introduce two-dimensional analogues of the latter, called *spin foams*.[80] Basically, a spin foam is a two-dimensional complex built from vertices, edges, and polygonal faces, with the faces labelled by group representations, and the edges labelled by intertwining operators. If we take a generic slice of a spin foam, we get a spin network. The first explication of spin foams Baez performs with a view to BF theory (as a first simple step).

For this, choose as a gauge group any Lie group G whose Lie algebra is equipped with an invariant non-degenerate bilinear form. Take as space-time any n-dimensional oriented smooth manifold M, and choose a principal G-bundle P over M. The basic fields are the connection A on P, an ad(P)-valued (n-2)-form E on M, where ad(P) refers to a vector bundle associated with P via the adjoint action of G onto its Lie algebra, and the curvature of A, which is an ad(P)-valued 2-form F on M. The Lagrangian for BF theory is simply $L = \mathrm{tr}\,(E \wedge F)$. Setting the variation of the action zero gives the field equations: $\delta \int_M L = 0 \Rightarrow F = 0$, $d_A E = 0$. Here, d is the exterior covariant derivative. The first equation tells us that the connection A is flat. BF theory shows up as a *topological field theory*, and locally all solutions look the same, because it does not have any local degrees of freedom. The second equation determines the gauge symmetries. The configuration space of BF theory is the space A of connections on P.[81] The corresponding classical phase space is the cotangent bundle T*A . Be $A_0$ the moduli space of flat connections on P, the physical phase space, and be G the group of gauge transformations of the bundle P. Then the *canonical quantization program* can be visualized by the following diagram:

---

$$T^*(\text{A }) \rightarrow \text{quantize} \rightarrow L^2(\text{A })$$
$$\downarrow \qquad\qquad\qquad \downarrow$$
$$\text{constrain} \qquad\qquad \text{constrain}$$
$$\downarrow \qquad\qquad\qquad \downarrow$$
$$T^*(\text{A }/\text{G}) \rightarrow \text{quantize} \rightarrow L^2(\text{A }/\text{G})$$
$$\downarrow \qquad\qquad\qquad \downarrow$$
$$\text{constrain} \qquad\qquad \text{constrain}$$
$$\downarrow \qquad\qquad\qquad \downarrow$$
$$T^*(\text{A}_0/\text{G}) \rightarrow \text{quantize} \rightarrow L^2(\text{A}_0/\text{G})$$

The problem is that typically, A  and A /G, are infinite-dimensional, making it difficult to define the related $L^2$-spaces.[82] But, as Baez points out, the great achievement of loop quantum gravity is that it gives background-free definitions of these Hilbert spaces by leaving the traditional Fock space formalism and taking holonomies along paths instead as basic variables. Hence, the basic excitations are not particles anymore (0-dimensional entities), but 1-dimensional spin networks. It turns out, in fact, that $L^2(\text{A }/\text{G})$ is actually being spanned by spin network states. Call such a state $\Psi \in$ Fun (A /G) so that any spin network in S ( = space) defines such a function. Because Fun is an algebra (namely consisting of all functions on A of the form $\Psi(A) = f(T \exp \int_{\gamma 1} A \ldots T \exp \int_{\gamma n} A)$, where f is a continuous complex-valued function of finitely many holonomies which are represented here by the integral expressions), multiplication by $\Psi$ defines an operator on Fun. We call this operator *spin network observable*. (It actually extends to a bounded operator on $L^2$, because $\Psi$ is bounded.) In fact, any product of Wilson loop observables can be written as a finite linear combination of spin network observables. Hence, the latter can be used to measure correlations among holonomies of A around a collection of loops. Moreover, it can be shown that a spin network edge labelled by the spin j contributes a length $(j(j + 1))^{1/2}$ (s.a.) to any curve it crosses transversely. Hence, length has a discrete spectrum of possible values in quantum gravity. Note that there is a very promising aside concerning triangulations: Given (n-1)-dimensional space S and any triangulation of it, choose a graph called „dual 1-skeleton", having one vertex at the centre of each (n-1)-simplex and one edge intersecting each (n-2)-simplex. We can express now any state in Fun(A$_0$/G) as a linear combination of states *coming from spin networks* whose underlying graph is this dual 1-skeleton. Using the holonomy picture we already know, we can define Hilbert spaces $L^2(\text{A}_\gamma)$ and $L^2(\text{A}_\gamma/\text{G}_\gamma)$ as before such that any spin network with $\gamma$ as its underlying graph defines a function $\Psi \in L^2(\text{A}_\gamma/\text{G}_\gamma)$. Spin network states are functions of the type $\Psi(A)$. They span the respective $L^2$-space, and we can therefore choose an orthonormal basis of them. Now, if $\gamma$ is a graph in S, trivializing the principal G-bundle with which S is equipped at the vertices of $\gamma$, gives a map A $\rightarrow$ A$_\gamma$ and a homomorphism G $\rightarrow$ G$_\gamma$ such that $L^2$ (A$_\gamma$) $\subset\rightarrow L^2$ (A) and the same

---

for A/G, $\hookrightarrow$ referring here to the inclusion mapping. When $\gamma$ is the dual 1-skeleton of a triangulation of S, for 3-dimensional Riemannian quantum gravity, it is always trivalent and particularly easy to visualize (having actually a hexagonal network pattern superimposing the triangulation). Spin numbers specify the lengths of the edges, with spin j corresponding to length $(j(j + 1))^{1/2}$. The same can be performed for 4-dimensional BF theory using 4-valent graphs now and tetrahedra.[83]

The definition of a *spin foam* now is very much alike the one for a spin network, only one dimension higher. A spin foam is essentially taking one spin network into another, of the form F: $\Psi \rightarrow \Psi$'. Just as spin networks are designed to merge the concepts of quantum state and geometry of space, spin foams shall serve the merging of concepts of quantum history and geometry of space-time.[84] Very much like Feynman diagrams do, also spin foams can be used to evaluate information about the history of a transition of which the amplitude is being determined. Hence, if $\Psi$ and $\Psi$' are spin networks with underlying graphs $\gamma$ and $\gamma$', then any spin foam F: $\Psi \rightarrow \Psi$' determines an operator from $L^2(A_\gamma /G_\gamma)$ to $L^2(A_\gamma /G_\gamma)$ denoted by O such that

$$<\Phi', O\ \Phi> = <\Phi', \Psi'><\Psi, \Phi>$$

for any states $\Phi$, $\Phi$'. The evolution operator Z(M) is a linear combination of these operators weighted by the amplitudes Z(O). Obviously, we can define a category with spin networks as objects and spin foams as morphisms.

So what we essentially do is the following: Given the (n-1)-dimensional S and a triangulation of S, choose a graph called the dual 1-skeleton. Express any state in Fun as a linear combination of states coming from spin networks whose underlying graph is this dual 1-skeleton. Define now space-time as a compact oriented cobordism M: S $\rightarrow$ S', where S, S' are compact oriented manifolds of dimension n-1. (Recall that two closed (n-1)-manifolds X and Y are said to be cobordant, if there is an n-manifold Z with boundary such that $\partial Z$ is the disjoint union of X and Y.[85]) Choose a triangulation of M such that the triangulations of S, S' with dual 1-skeletons $\gamma$, $\gamma$' can be determined. The basis for gauge-invariant Hilbert spaces is given by the respective spin networks. Then the evolution operator Z(M): $L^2(A_\gamma/G_\gamma) \rightarrow L^2(A_\gamma/G_\gamma)$ determines transition amplitudes $<\Psi'$, Z(M) $\Psi>$ with $\Psi$, $\Psi$' being spin network states. Write the amplitude as a sum over spin foams from $\Psi$ to $\Psi$': $< , > = \sum_{F:\Psi \rightarrow \Psi'}$ Z(F) plus composition rules such that Z(F') o Z(F) = Z(F' o F). This is a discrete version of a path integral. Hence, re-arrangement of spin numbers on the „combinatorial level" is equivalent to an evolution of states in terms of Hilbert spaces in the „quantum picture" and effectively changes the topology of space on the „cobordism level". This can be understood as a kind of *manifold morphogenesis* in time: Visualize the n-dimensional manifold M (with ∂M = S ∪ S' - disjointly) as M: S → S', that is as a process (or as time) passing from S to S'. This the mentioned *cobordism*. Note that composition of cobordisms holds and is associative, but *not commutative*. The identity cobordism can be interpreted as describing a passage of time when topology stays constant. If there is no change of topology (due to the action of the identity cobordism), then there is no change of state, because we do not have any local degrees of freedom here. Visualized this way, TQFT might suggest that general relativity and quantum theory are not so different after all. In fact*, the concepts of space and state turn out to be two aspects of a unified whole, likewise space-time and process*.

Note that „time" shows up here not as a function, but as a manifold (although arrows are used). This is particularly interesting, because with a view to what Barbour tells us about the „absence" of time, this means that the concept of time is intrinsically included here as a pragmatic ordering principle for localizing topology changes. This is similar to what Prigogine calls the „age of a system", which is roughly a frequency of formations of new structures in a system making the latter more complex. Time as a convention then, would be an approximate „average" over such ages. (I have commented on this in more detail at another place.[86]) Hence, time shows up as being associated to a kind of measuring device for local complexity gradients. So what we have in the end, is a rough (and simplified) outline of the foundations of emergence, in the sense that we can localize the fine structure of emergence (the re-arrangements of spin numbers in purely combinatorial terms being visualized as a motion-in-itself) and its results on the „macroscopic" scale (as a change of topology being visualized by physical observers as a motion-for-itself). This is actually what we would expect of a proper theory of emergence. But note also that space and time, in the classical sense, are obviously absent on a fundamental level of the theory, but can be recovered as concepts when tracing the way „upward" to macroscopic structures. In other words: even as a gross average feature for „shortsighted" human scientists (as Penrose indicates it at the end of his first twistor paper), space and time would nevertheless turn up as (philosophical) categories of concepts, simply, because the meaning of these categories is well-adapted to what humans actually perceive of their world (and communicate to other humans). This is in fact, a point, where Barbour's argument seems to break down (if discussed within this philosophical perspective): What he essentially shows in his book is that quantum theory, *in so far as it is foundational*, describes partly what was called non-being (or substance) in former times. Hence, there is neither space nor time in *real* terms ( = realiter, i.e. with respect to what there is in an absolute sense of the world's foundation), but there *is* space and time in *modal* terms ( = modaliter, i.e. with respect to what humans perceive of their world). The former refers to substance, or, alternatively, to what

---

[86] R.E.Zimmermann: Selbstreferenz und poetische Praxis, Junghans, Cuxhaven, 1991.

*speculative* philosophy is all about. The latter refers to the physical world, or, to what *sceptical* philosophy is all about. The one relies on theoretical speculation according to what we know - speculating about the foundation of the world, which is outside (logically „before") the world, and of which we are not a part therefore, and hence, about which we cannot actually know anything. The other refers to the empirical world, about which, with the help of experiments, we can obtain knowledge, in fact. Obviously, in terms of physics, the first (speculative) aspect is corresponding to physical theory, in so far as it is foundational. The second (sceptical) aspect corresponds to physical theory, in so far as it is empirical.[87]

These results can also be formulated in the language of category theory: As TQFT maps each manifold S representing space to a Hilbert space Z(S) and each cobordism M: S → S' representing space-time to an operator Z(M): Z(S) → Z(S') such that composition and identities are preserved, this means that TQFT is a functor Z: nCob → Hilb. Note that the non-commutativity of operators in quantum theory corresponds to non-commutativity of composing cobordisms, and adjoint operation in quantum theory turning an operator A: H → H' into A*: H' → H corresponds to the operation of reversing the roles of past and future in a cobordism M: S → S' obtaining M*: S' → S.[88]

## 5  Topoi - Wordly & Pre-Wordly

We have already noted the significance of categories for the approaches we are interested in. Louis Crane has announced that category theory would probably be a unifying principle in physics.[89] But this is not simply a point of formal conceptualization. It is also an important aspect of the process of thinking itself. For the first time, Vladimir Trifonov has made this aspect explicit, as a consequence of applying topos-theoretic concepts to physics.[90] He introduces topoi (toposes)[91] as abstract worlds which represent universes of mathematical discourse whose inhabitants can utilize non-Boolean logics for their argumentation (i.e. their propositional structures). In contrary to the *sensory space* which mainly describes the observations of observers ( = researchers), the space of motions is the *set of actions*. Be F a partially ordered field. Then, an F-*xenomorph* is a category A(F) of linear algebras over F. The objects of A(F) are called *paradigms* of an F-xenomorph, the arrows are called *actions*. Essentially, a paradigm then, is the set of states of knowledge. A paradigm is called *rational*, if the space of motions M(A) is a monoid. In particular, it can be shown that the set of all possible actions of a researcher is a topos whose arrows are those mappings which preserve realizations of the monoid (of the space of motions). It can also be shown that, if A is a rational paradigm, and the topos of all possible actions is Boolean (non-Boolean), then the paradigm A is classical (non-classical). For a xenomorph F = R of a generic type of the observer's psychology, Trifonov can finally show that an R-xenomorph implies a classical Einstein paradigm, i.e. dimension 4 and signature 2 of the space-time metric. Also: If A is a non-trivial Grassmann algebra, then the paradigm is the Grassmannian of an R-xenomorph. Because A has a zero divisor, M(A) cannot be a group. Hence, the logic of a Grassmannian paradigm is always non-Boolean, and the mathematics is non-classical. As I have discussed at another place[92], it is very likely that the category of negators (essentially operators acting upon „world states" in order to produce complexity, which can also be visualized as chaotic self-compositions of some suitable, unfolding „ground state" of the world) forms a topos. (They may even turn out to be basically identical with the functor Z: nCob → Hilb, discussed in the preceding section in terms of TQFT.) The interesting point in the conception of Trifonov's is that (the logically formal part of) *thinking itself* is directly related to the physical process of unfolding the worldly structure as it can be described in terms of cosmological evolution. The basic idea in this is to define a self-referent cycle in the sense that the physical process is producing observers who choose their explicit logic for evaluating what they actually observe. We recognize the idea again, of nature exploring itself by means of human research (or telling its own story to itself, as a kind of self-narration which is modelling its own self-unfolding).

---

[87] Do not think that both things would be the same for physics: In general relativity, we can clearly recognize that there is a well-defined conceptual part (in other words: a foundational part) which is actually clarifying the attitude of the approach, but without giving a pathway towards empirical results. But when the Einstein field equations are being introduced, then testable hypotheses can be formulated and checked in observational experiments. Probably, the same is true for quantum theory: In so far it is foundational, one would not expect any directions for experiments. Hence, Barbour would not have to defend his approach against Fay Dowker's objection of not having to offer any testable prediction. A fundamental theory is not actually obliged to make such an offer. Essentially, Barbour is telling us that there is no time at a fundamental level. But then, this is not a revolution at all, because in physics we learnt this already within the development of our century. In philosophy, already Spinoza has formulated this quite clearly, a long time ago.

[88] For a somewhat different approach to spin network evolution see also F.Markopoulou, L.Smolin: Causal Evolution of Spin Networks, Nucl.Phys. B 508, 1997, 409, and id.: Quantum Geometry with Intrinsic Local Causality, gr-qc 9712067 (16/12/77). See also more recently F.Markopoulou: Dual Formulation of Spin Network Evolution, gr-qc 9704013, id.: The Internal Logic of Causal Sets: What the Universe Looks Like from the Inside, gr-qc 9811053, and also id.: Quantum Causal Histories, hep-th 9904009 v4 (4/6/99).

[89] As quoted according to J.C.Baez: This Week's Finds in Mathematical Physics, op.cit., week 31, p.1.

---

Note however that this time, we have a sound mathematical base for giving a consistent foundation for such a view. In fact, the speculative part of the theory (as far as the aspect of substance and its relationship to its own attributes is concerned) is at least partially formalized, with respect to two important points: On the one hand, it is formalized in being integrated into the description of the physical process itself. On the other hand, it is formalized according to the process of thinking such as to give an explicit choice of logical types which can be utilized for interpreting observations made. The crucial aspect here is in particular that the appropriate logical type can be used to reproduce the phenomenological structure of the worldly observations actually being made (as we know them). Take irreversibility for instance: Types of logic which have a modified law of negation, of the kind $\neg (\neg x) \neq x$, if x is a given proposition, determine the phenomenology as it is actually being observed according to the fact that processes are irreversible in the sense that their logical representation cannot reproduce initial propositions, independent of the number of negation operations acting repeatedly on such propositions. In other words: Recursive operations of this type have no fixed points. In a sense, we can say that temporality is coming in explicitly where earlier the logic remained static all the time (and created considerable difficulties when comparing theory with praxis, as e.g. Lacan has shown in some detail[93]). Hence, the advantage of topoi: They operate in terms of an intrinsic concept of time which can be visualized as a kind of generic concept, unifying the object level of a theory (that about which the theory is actually speaking) with the subject level (which determines the logic of the observer who actually speaks - in terms of the theory). This is another indication as to the phenomenological necessity of time on a macroscopic level of worldly perception and reflexion. (But, as said before, this does not alter the fact that on a fundamental level, time as a concept, may be absent altogether.)[94]

To this end, we note that the process of the concrete, physical unfolding of the world (as it can be assessed in empirical terms) is essentially identical with the process of reflecting about it, in a cyclic manner which secures that the egg comes before the hen. Following Sandkühler here[95], we call this aspect „onto-epistemic“, in the sense that both the ontological and epistemological components of the human mode of grasping the world operate very much on the same footing (also very close to Spinoza's argument given in his 2p7, as we have seen earlier).

But in the meantime, we can do a little more, thanks to recent research undertaken in terms of topos theory. We will shortly elaborate on this in the following. We begin with recent work of Isham and Butterfield on topos theory and quantum physics.[96]

This ansatz is particularly interesting, because it starts from a propositional viewpoint, in first place: Basically, the Kochen-Specker theorem states the impossibility of assigning values to all physical quantities, when (at the same time) preserving the functional relations among them. Or, in more technical terms: If dim H is greater than two, there are no global valuations. Reconceive of a valuation now, as giving truth values to *propositions about the values* of a physical quantity rather than assigning a value to the quantity itself. Here, a certain amount of contextuality is involved, in which a value ascribed to a quantity cannot be part of a global assignment of values, but must instead depend on some context. The idea is then, to eventually re-introduce globality, but for the price of ending up with *partial* truth values, which means that the truth value of a proposition belongs to a logical structure that is larger than $\{0,1\}$, and these target-logics are context-dependent. In particular, the space of contexts is the category of all Boolean subalgebras of the projection lattice (rather than the category of self-adjoint operators).

There are two basic aspects to this ansatz: The *first* one is that a theorem can be proven[97] which states that to each generalized valuation v, there is a natural transformation $V: \Sigma \rightarrow \Omega$ for which, at each stage of truth, the component with respect to ⓒA, is defined by $V_A(a) := v(A = a)$. Here, ⓒA refers to the spectral representation of bounded self-adjoint operators on a Hilbert space H. They are the objects of a category O for which the morphisms are maps ⓒB $\rightarrow$ ⓒA, if there is a Borel function f: $\sigma$(ⓒA) $\rightarrow$ R such that ⓒB = f(ⓒA), when $\sigma$ is the spectrum, and ⓒA = $\int_\sigma \lambda$ dE, with E being the spectral projection operators. Then, *a generalized valuation on propositions* in a quantum theory is a map v that associates with each group of the form A∈$\Delta$, $\Delta$ being a Borel subset of $\sigma$(ⓒA), a sieve v(A∈$\Delta$) on ⓒA in O. This is actually the crucial point: to associate a sieve, which is the reason for $\Omega$ in the theorem to be a subobject classifier.

Recall that in fact, parallel to set theory, where subsets of some set are in 1-1 correspondence with characteristic functions whose target space can be visualized as giving the simplest „false-true"-Boolean algebra, in topos theory, subobjects of an object are similarly in 1-1 correspondence with „characteristic" morphisms having a special object, called the *subobject classifier*, as their target space which is analogous to $\{0,1\}$. Take a poset C then: A function that assigns to each p∈C a set $X_p$, and to each pair p ≤ q a map $X_{qp}$: $X_q \rightarrow X_p$ such that $X_{pp}$ = id ($X_p$), and whenever p ≤ q ≤ r, then $X_{rp} = X_{qp}$ o $X_{rq}$, is called a *pre-sheaf* X on C. A *subob-*

---

*ject* K of a pre-sheaf X is then essentially another pre-sheaf with a similar map, $K_{qp}$ say, which is the restriction of $X_{qp}$. The collection of all pre-sheaves on a poset C forms a category, denoted $Set^C$. This can be shown to be a topos then, because pre-sheaves can be defined via *sieves*, which are collections of morphisms acting on objects of C such that compositions are preserved. The crucial property of sieves is that there are subobject classifiers which have the structure of a *Heyting algebra*. (To see this note with the above that, given a pre-sheaf $\Omega$: C → Set, and an object A of C, then $\Omega(A)$ is the set of all sieves on A, and if f: B → A, then $\Omega(f)$: $\Omega(A)$ → $\Omega(B)$ is defined as the pull-back to B of the sieve S on A by the morphism f:

$$\Omega(f)(S) := \{h: C \to B \mid f \circ h \in S\},$$

for all $S \in \Omega(A)$.) The existence of a subobject classifier can be taken as a defining property of a topos. The former does turn out to be an object of possible truth values such that there is a characteristic morphism in the above mentioned sense, which, at each „stage of truth" A in C, can be written like

$$\chi_A(x) := \{f: B \to A \mid X(f)(x) \in K(B)\},$$

for all $x \in X(A)$. Hence, each stage of truth serves as a possible context for an assignment to each proposition x of a generalized truth value, which is a sieve belonging to the Heyting algebra $\Omega$. (This being the result of finding out that a valuation on propositions must be some sort of structure-preserving function from the set of propositions to the set of truth values of logical algebra. Within this language, the Kochen-Specker theorem can be rephrased, saying that there is no global section of pre-sheaves arising in quantum theory. Defining generalized valuations then, whose values are sieves of operators, can be used to show that each quantum state actually generates such a valuation.)

The basic ideas for this go back to Lawvere who in the late seventies developed a very elegant concept of motion as derived from logical properties of topoi.[98] As a short aside note that in particular, the states X of a body B should be sufficient to determine their own evolution provided the general law of motion L (in the Lagrangian sense) is known. In fact, X may involve histories of motion, but there is always a morphism X → Q (the configuration space of B), expressing the fact that each state involves a specific underlying configuration. Classically, the state space will be X = $Q^D$ (the tangent bundle) meaning that infinitesimal histories are all that is necessary. The configuration space Q ⊂→ $E^B$ (subspace of admissible placements) implies that $X \subset\longrightarrow Q \times V^B$ (space of velocity fields on B). The dynamically possible motions of B can be singled out from the kinetically possible ($Q^T$) by knowing the Lagrangian ʟ: $X \to W$, where the latter refers to the work needed to add to the potential energy of q in order to obtain the kinetic energy of the velocity field v. Note that q a morphism T x B $\to$ E, describing the motion of B in E, where E is usually equipped with a Euklidean metric d: E x E $\to$ R. Then, the action is given by

$$I = \int ʟ \, (q, \partial q / \partial t) \, dt,$$

which is a morphism $Q \to A = W \otimes T$, $Q \subset\longrightarrow Q^T$. The object of the ʟ-possible motions is then the subobject of $Q$ such that grad (I) vanishes. This gives the usual Lagrangian equations of motions.[99] To the aspect of having a concept of „intrinsic motion" we will come back later again.

The *second* aspect in Isham and Butterfield is more in the propositional field[100]. Although discussed in somewhat intuitive terms, the explication that truth values are essentially sieves, turns out to be of a significant relevance: Given a category C, and to each object $A \in C$, a set P(A). For each A and $d \in P(A)$, [A,d] shall be thought of as a proposition. If there is a morphism f: B $\to$ A, then there is also a function f*: {[A,d] │ $d \in P(A)$} $\to$ {[B,e] │ $e \in P(B)$} acting on the d's. Hence, given f, then [B, f*(d)] is the B-proposition that corresponds to the respective A-proposition by f. Note that if now the composition f: B $\to$ A, g: C $\to$ B; f o g: C $\to$ A with g*(f*(d)) = (F o g)*(d) is *not* satisfied, then the *-operation is actually *path-dependent*: If a morphism k: C $\to$ A can be factored as C $\to_g$B $\to_f$A, then the pull-back k*(d) of $d \in P(A)$ *may not be equal to* g*(f*(d)) obtained by factoring through B. While the authors refer to this situation in physical terms as being clearly „pathological", it may be exactly this case which is common for most hermeneutic systems of propositional structures. We will come back to that in the final section.

The general idea is then to assign truth values to each of the propositions. Obviously, [B, f*(d)] is logically weaker than [A,d], because it is its consequence ($\leq$). If now v(A,d) is the truth value assigned to [A,d], then if the latter is totally true, so are all of its consequences. If it is partially true, it is more true the more its consequences are totally true. The truth value v(A,d) is to be determined by which of the consequences [B, f*(d)] of [A,d] is totally true. So there is actually the possibility to define „truth distances", and the semantic value (contents) of a proposition is actually being determined by the set of those of its consequences that are true. Hence, v(A,d) is the set of morphisms f: B $\to$ A such that the associated [B, f*(d)] is totally true. (Recall that „totally" and „partially" refer to the cases according to whether the characteristic object is {0,1} only, or larger than that, respectively.) The proposal is then, that for any [A,d], total truth is just v(A,d) being the set of all morphisms of type f (which is actually the principal sieve on A) - underlying the idea that v(A,d) is a sieve, indeed.

Obviously, these aspects of propositional semantics are of some importance within the field of computer science. Although computer logic might turn out to be not equal to human logic, it is nevertheless very interesting to have a look for possible applications of the above said to this domain.

So, some time ago, starting from a somewhat different perspective, Abramsky and Vickers have begun to place notions of *observing and testing processes* within an algebraic framework in which observations effectively constitute a quantale, and the propositions of geometric logic are related to the logic of finitely observable properties.[101] To this purpose they define *topological systems* to be essentially topological spaces and locales, at the same time, the latter being homomorphisms from the frames of open sets (the complete Heyting algebras) to the set 2. The idea is then that finitely observable properties closed under propositional connectives of geometric logic only, give a computational interpretation of topology and domain theory in some logical form.

A *quantale* is basically a sup-lattice (a complete join semilattice) equipped with a monoid structure satisfying distributive laws. Hence, quantales are to linear logic as frames are to intuitionistic logic.[102] The programme is then, to use the algebraic framework of modules over quantales to analyze a process equivalence. One of the interesting results is the following: Take typed semantics with two types which are objects of a *quantaloid*. This is a small category such that each homomorphism set is a sup-lattice, and the morphism composition distributes over all joins. (In fact, Hom(quantaloid) is a functor that preserves all joins.) The types are called live (*) and dead ($\perp$). Actions are observed of a live process, refusals and acceptances are postmortem observations. Introduce ♠ to mark the transition from life to death. Then we get a directed graph which generates the *ready* quantaloid:

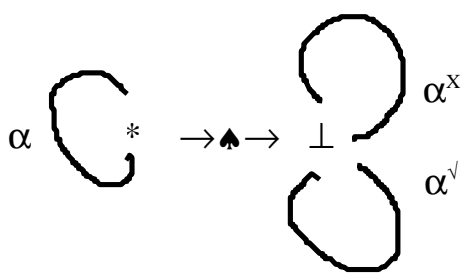

What does that actually mean? Basically, the idea deals with the testing of equivalences of processes, defining the respective types of equivalence by quantales with a testing preorder, using Act as the set of atomic actions. Then, a *transition system* labelled over Act is a set Proc equipped with a transition relation $\to \subseteq$ Proc x Act x Proc. Quantales can be shown to actually generalize both topological spaces and transition systems. Technical refinements of this[103] do yield a whole family of testing equivalences on processes, as instances of an algebraically formulated axiomatic framework. The diagram deals actually with three of them: Ready (R), Failure (F), and Acceptance (A), which are closely related to their respective traces RT, FT, and AT, but with the property that after refusal or acceptance, no more pure actions are possible. They are represented by a suitable quantaloid, and according to a suggestion of Abramsky[104], that different process equivalences represent equivalence in behaviour under different notions of how the processes can be tested or observed, certain fundamental observations are formalized as generators of quantales. They are called subbasic (in analogy with topology), and are defined with respect to a fixed set Act of process actions. Hence, $\alpha$ is the observation that action $\alpha$ has been effectively performed, along with any associated change of state. Then $\alpha^x$ is the *refusal* of $\alpha$, meaning the observation that the process has signalled its inability to perform $\alpha$. There is no change of state (although the state of knowledge improves). Finally, $\alpha^{\sqrt{}}$ is the *acceptance* of $\alpha$, i.e. the observation that the process has signalled its ability to perform $\alpha$, although it has not yet done so. The latter two are propositional in nature.

Take e.g. the failures semantics F: Let Act be a set. Then we may present the quantaloid of the form

$$Q = \{\alpha\colon * \to * \ (\alpha \in \text{Act}), \ \spadesuit\colon * \to \bot, \ \alpha^x\colon \bot \to \bot \mid$$
$$\alpha^x \bullet \alpha^x \le 1_\bot, \ \alpha^x \bullet \beta^x = \beta^x \bullet \alpha^x\}$$

and the testing order by

$$1_* \le \spadesuit,$$
$$\spadesuit \bullet X^x \le \alpha \vee \spadesuit \bullet (X \cup \{\alpha\})^x.$$

The first inequality tells us that if a process is live, then it can die. For the second one, suppose that p is a live process, and that after death, a postmortem examination reveals that it would refuse the actions in X: $p \bullet \spadesuit \bullet X^x \ne 0$. Consider then, whether p could have done $\alpha$. If so, then $p \bullet \alpha \ne 0$; if not, then a more careful postmortem examination would reveal that p would refuse the actions in $X \cup \{\alpha\}$: $p \bullet \spadesuit \bullet (X \cup \{\alpha\})^X \neq 0$. Inherent in this is the notion of the meaning of a process: The *meaning of a process* is given by the set of its capabilities: Be $a \in Q$ a quantale such that $\{p\} \bullet a \neq \emptyset$. Then construct the semantic domain for processes out of Q. The question is now when two elements of Q might be equivalent as process capabilities: Given a, b in Q, when do we have that for every process p, $\{p\} \bullet a \neq \emptyset$ iff $\{p\} \bullet b \neq \emptyset$? This is basically what this kind of „algebraic semantics" is all about.

# 6 Preliminary Conclusions

We collect now the most important results from the last two sections and relate them to what we have said in the first two sections as to the underlying intention of our enterprise. We will go backward for this, and take the last aspects first: Hence, as the initial concept, the *meaning of a process* following its observation shows up, especially with a view to classifying equivalent processes visualized as operations on quantities which in turn classify transitions (between states of systems). This may be interpreted as the actual beginning of the process of scientific research which is also the process of human reflexion (in particular: of reflecting about nature). The formalization of this in terms of typed semantics (leading up to the definition of quantaloids) means basically that reflexion has to rely on the abstract mapping of its results, by including them within an algebraic (or topological, or geometric) framework. Obviously, this is due to semiological conditions which are boundary conditions for human reflexion. On the other hand, these boundary conditions are self-imposed by the cognitive system of humans, because they simply express the human capabilities of „working with" what is perceived. Hence, the close relationship between algebraic, topological, and geometric structures on the one hand, and logic on the other. The relevance of Trifonov's work lies in the fact that it can be shown that formally, the actual choice of human logic (for the practise of reflexions) is nothing but a product of the process which is the object of research in terms of this very logic. The relevance of the work of Abramsky and Vickers is to show that in terms of topos theory, this logic can be explicitly related to the (mathematical) structures mentioned above. (And in fact, for a quite generalized framework, because we would expect that typically, the computer logic (of simulations e.g.) turns out to be different from human logic.)[105]

---

[105] For the characteristic relationships of algebraic and geometric theories within the framework of topoi see P.T.Johnstone: Topos Theory, Academic Press, London etc., 1977. - I have discussed the semiological aspects of this, and in particular, the meaning of metaphors with a view to theories, in more detail in my „The Klymene Principle", op.cit., section III D. - In his „Toposes *pour les nuls*", Steve Vickers, starting from categories of sheaves as generalized universes of sets, comments on geometric constructions, and refers this more to model theoretic aspects. He shows in which sense a classifying topos can be interpreted as a space of models of a geometric theory.

Hence, categories, and especially topoi, provide a formalized (mathematical) structure for computing purposes, for the modelling of processes, and for propositional structures (of modelling the modelling), at the same time. This is the chief advantage of categories: that they make explicit *the onto-epistemic character* of research (i.e. reflexion). Moreover, the intrinsic logic shows that observers who interpret their world in a non-Boolean manner, are of generic type (or: may be „more generic" than Boolean observers).

Reverse the order of argument now: When we think of humans with a logic as products of nature with a physics, then the choice of logic is actually *an outcome* of the physical process, in first place, rather than something which is imposed upon nature by some „external" (or: independent) physical (human) observer. This is a somewhat stronger rephrasing of Trifonov's argument, in fact. The propositional aspect as derived in the paper by Isham and Butterfield then, deals mainly with the problem of mapping theory-languages onto each other: The point is to actually discuss „translational" aspects of semantics. It is important now to notice the role of the composition rules: If they are path-independent operations (acting on propositions), they refer to what Trifonov calls a *rational paradigm* of logic. If however, they are path-dependent, they refer to a *hermeneutic paradigm* instead.[106] Hence, the latter can be referred to the production of excess meanings, which are characteristic for „non-formalized" situations.

This can also be discussed in more physical terms: In fact, for Vickers (though he mentions this within another context), the category of sheaves is essentially a generalized universe of sets. In this sense, Trifonov speaks of „abstract worlds" which are „universes of mathematical discourse". But, on the other hand, given any category C, the pre-sheaf on C is a contravariant functor F: C $\rightarrow$ Set. If, in particular, C is the category of shapes, then the morphisms correspond to ways the shapes can be glued together to give one shape (or to be more precise: there are morphisms of the type f: x $\rightarrow$ y such that x can be included one way or the other in y). Then a pre-sheaf on C can be thought of as a geometric structure built by gluing together these shapes along their common pieces. (Similarly, a pre-sheaf on the category of simplices turns out to be a „kind of space".)[107] Note that there is a choice of configurations in a way, because when we introduce a special pre-sheaf, the subobject classifier, then we actually deal with generating truth values (in the sense of Isham and Butterfield). In particular, we can think of TQFT as a pre-sheaf of Hilbert spaces on the category nCob whose morphisms are n-dimensional cobordisms. Hence, TQFT is a Hilbert space object in the topos of pre-sheaves on nCob. So, it is a quantum theory, because of it's being a Hilbert space object, while its peculiar variability (in assigning different Hilbert spaces to each (n-1)-dimensional manifold representing space) is due to it's being an object in a topos, because this property is expressing the *aspect of intrinsic motion* ( = change/variation).

---

[106] Note that we would speak of „hermeneutic" rather than of „irrational", because although there is a good deal of irrationality in what is at stake in hermeneutics, the „space of free play" associated with a more flexible amount of „excess meaning" typical for hermeneutic situations, can also be attributed to a mere lack of information under simultaneously performed rational reflexion.

[107] This example is mentioned in J.C.Baez, This Week's Finds ..., op.cit., week 115.

Hence, we can clearly recognize the close interrelationship between the physical states and their representation in terms of „form" (morphology) on the one hand, and their logical conceptualization on the other. This does not actually mean however, to explicitly *derive* logic from physics. But it may demonstrate (for a start) both the intrinsic equivalence and the strict parallelism of the two, very much in the sense of our modern re-interpretation of Spinoza's original intention. And there are many examples by now, which can illustrate the general tendency of what is actually being done: Note e.g. Kauffman's remarks on knot theory and DNA, clearly displaying their semiological relevance.[108] See also the work of Casati and Varzi who give an explicit history of discontinuities[109] modelled according to a parallel with macroscopic morphogenesis. Although the authors do not refer to it, there is an obvious analogy with the perceptive and cognitive processes which are at the basis of human reflexion, as Patricia Churchland has discussed them.[110] On the lowest (or highest, but in anyway most fundamental) level then, we have spin foams which turn out to be equivalent to a „microscopic" picture of an evolution operator acting on Hilbert spaces, determining transition amplitudes of spin network states. Note the abstract character of this underlying foundation of the world, from which physical structures and forms of matter are eventually emerging (the human mind being one of them). Hence, the connotation of substance. Note also that *production* means here differentiation rather: The world can be visualized as a *self-differentiation* of the ground, gradually unfolding the complete hierarchy of worldly structures.

In order to represent the hierarchy so achieved we take „really important" categories and express them in terms of Chinese (!) characters: We start with the foundation of all, given by the abstract structures of spin networks and spin foams, or, alternatively, by Hilbert space operations, going forward to concrete cobordisms, producing changes of topology, then to the geometric picture implying loops and connections, and curvature (or forces) in the end. This is the physical level of

---

[108] L.H.Kaufmann: Knots and Physics, op.cit., 423sqq. - Here molecular processes, knot theoretic operations, and the toplogical information of knots are closely connected to each other. Hence, the DNA shows up as the fundamental carrier of morphogenetic fields: the knotted form functioning as a receiver of field information, knots forming a kind of alphabet of the field language, generating a lexicographic order. This is actually very much in the sense of what Thom expected as a result coming from the semiological interpretation of his catastrophe theory. Cf. my „René Thom - Semiologie des Chaos", in: G.Abel (ed.), Französische Nachkriegsphilosophie, Autoren und Positionen, in press. Note also that according to Rovelli and Smolin, loop states are in 1-1 correspondance with knots in space, and that spin networks (lying at the basis of loops) are simplicial quantum gravity (Haslacher, Perry, 1981). Hence, the relationship between the latter and morphogenetic information. This might be the key for solving the „genetic problem" of Smolin's theory of cosmological natural selection, ene Principle", ch. VI.

[109] R.Casati, A.C.Varzi: Holes, MIT Press, 1994.

[110] P.S.Churchland: Neurophilosophy, Toward a Unified Science of the Mind/Brain, MIT Press, 1992 (1986). Note in particular: 428-447. We find here the remarkable formulation: „There is no reason to imply that the

worldly processes. At the same time, the fundamental level implies the level of reflexion in terms of logic and hermeneutic, the latter being also produced in the physical sense, but they themselves „acting back" again on what produced them, logic acting upon physics, in attempting to actually represent it, hermeneutic acting *on logic and physics*, respectively, in attempting to represent representations:

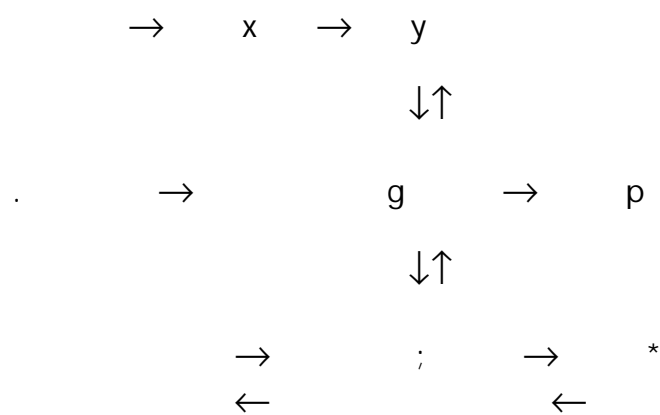

We start from the left-hand side with the foundation itself, which is pronounced *tchi* ( . ) and means „ground". The physical line of production is the upper one, beginning with *en* (cycle) ( x ), and producing *getsu* (sickle moon = curvature) ( y ) and hence concrete physics in macroscopic terms. The physical structures emerge from the left out of their foundation, but they themselves also produce one of their forms which is rational mind (logic), expressed by *bun* (sentence/proposition) ( g ). This in turn produces, through its spaces of free play, excess meanings, and thus hermeneutic, expressed by *wu'a* (harmony)( ; ). The one produces knowledge laid down in various disciplines, expressed by *gaku* (subject disciplines) ( p ). The other produces understanding and insight, expressed by *dan* (the rising sun) ( * ). It is the latter only which can act back onto the ground ( = foundation).

Independent of this somewhat metaphorical representation of a systematic approach to what we call *transcendental materialism* (because of obvious reasons), we may note that, in principle, what we have here is a category whose objects are categories, and whose morphisms are functors. Recall that „the theory of categories" itself is a category with finite limits whose models are categories. And in this case, models are functors to Set preserving finite limits.[111] If we could show this for each of our categories, functors of this kind would be *world models*, actually representing (and expressing) the foundation of the world. The interesting point is that part of the foundation itself (remember the spin foams) can be modelled in terms of (mathematical) physics. But in so far this is a part of the foun-

---

dation of the world rather than of the world itself, the physical theory associated to it, is a *foundational theory*, in the sense that it precludes empirical confirmation. (This is basically true for all conceptual theories of a unifying character.) Hence, it is the formalized counterpart of speculative philosophy. Whilst empirical theories which are subject to experimental testing describe the physics within the world. Unless they are not added to the foundational ingredients of a theory, the latter is not testable. It is not its objective to be testable in fact, because it is designed in order to represent worldly foundation. Hence, empirical theories are the formalized counterpart of sceptical philosophy. This relationship is essentially discussed here in terms of an analogy which draws its expressive potential from the classical Spinozist relationship between substance and attribute. Although today, we would not really think of Spinoza's philosophy as a solution to our present problems, it proves nevertheless useful to actually re-construct the latter within the context of a modernized version of the general framework Spinoza's theory laid down as a guideline for further orientation. This is in fact, what philosophy is all about. And in this sense, philosophy might have its own (heuristic) merits.

## Dedication

This work is dedicated to the memory of Alexandros Chapsiadis.
„O gebt euch der Natur, eh' sie euch nimmt."[112] (F.Hoelderlin)

## Acknowledgements

For illuminating discussions I would like to thank John Baez, Julian Barbour, Richard Bell, Mary Hesse, Chris Isham, David Robson, Lee Smolin, John Spudich, Steve Vickers, and Paola Zizzi. In particular, I thank Basil Hiley for his interest in this paper and the discussions we had about its implications, as well as for giving me the occasion to present parts of it to his theory group at Birkbeck College London.

---

[112] „Oh, give yourself to Nature, before she herself takes you." From: Der Tod des Empedokles, 1st version, HKSA (standard edition) IV, 69.